\newcommand\nodata{{...} }

\newcommand\etal{et~al.}
\newcommand\kms{\ifmmode {\rm\,km\,s^{-1}}\else${\rm\,km\,s^{-1}}$\fi}
\newcommand\simlt{\mathrel{\spose{\lower 3pt\hbox{$\mathchar"218$}} \raise 2.0pt\hbox{$\mathchar"13C$}}}
\newcommand\simgt{\mathrel{\spose{\lower 3pt\hbox{$\mathchar"218$}} \raise 2.0pt\hbox{$\mathchar"13E$}}}
\newcommand\aap{{\em A\&A}}
\newcommand\aasup{{A\&AS}}
\newcommand\aaps{{\em A\&AS}}
\newcommand\aj{{\em AJ}}
\newcommand\apj{{\em ApJ}}
\newcommand\apjl{{\em ApJ}}
\newcommand\apjs{{\em ApJS}}

\newcommand\mnras{{\em MNRAS}}
\newcommand\nat{{\em Nature}}

\documentclass[usenatbib]{mn2e}

\usepackage{psfig}
\usepackage{rotating}
\usepackage{amssymb}

\title[Clustering and properties of galaxies in distant radio galaxies]{
Clustering and properties of $K$--band companion galaxies around USS radio sources\thanks{Based on observations obtained with
the Australia Telescope Compact Array, the Anglo-Australian Telescope,
and the European Southern Observatory, La Silla, Chile (Program
70.A-0514)}.}
\author[C. Bornancini \etal]{
\parbox[t]{\textwidth}{
Carlos~G.~Bornancini$^1$, Diego~G.~Lambas$^{1,2}$
Carlos De Breuck$^3$.}
\vspace*{6pt} \\ 
$^1$Grupo de Investigaciones en Astronom\'\i a Te\'orica y Experimental, IATE\\
Observatorio Astron\'omico, Universidad Nacional de C\'ordoba\\
Laprida 854, X5000BGR, C\'ordoba, Argentina.\\
$^2$Consejo Nacional de Investigaciones Cient\'\i ficas y T\'ecnicas (CONICET),
Avenida Rivadavia 1917, C1033AAJ, Buenos Aires, Argentina.\\
$^3$ European Southern Observatory, Karl Schwarzschild Stra\ss e 2, D-85748 Garching, Germany.\\
}
\pubyear{2005}
\begin{document}
\maketitle

\begin{abstract}

We have analyzed galaxy properties in the environment of a sample of 70  
Ultra Steep Spectrum (USS) radio sources selected from the Sydney University Molonglo sky Survey 
and the NRAO VLA Sky Survey catalogues, using near-IR data complete down to $K_s=20$.
We have quantified galaxy excess around USS targets using an Abell--type measurement $N_{0.5}$ \citep{hill}.
We find that most of the USS fields studied are compatible with being Abell class 0 richness clusters.
A statistical analysis of the distribution of companion galaxies around USS radio sources show a 
pronounced tendency for such objects to be found in the direction defined by the radio axis, 
suggesting that they may be related to the presence of the radio sources.
We have also measure the central concentration of light of the USS sample and compare these to the values 
obtained for field galaxies and galaxies selected through other methods. 
By using Spearman statistics to disentangle richnesses and concentration indices dependences, we detect a weak, but significant, positive correlation.
We find that at $z>2$ USS radio sources are more concentrated than field galaxies at similar redshifts, indicating that these objects trace the most massive systems at high redshift.

\end{abstract}


\section{Introduction}
Hierarchical galaxy formation models predicts that large galaxies and other massive structures, grow from mergers of small subunit of mass. 
Studying the properties of distant clusters of galaxies and their evolution can directly constrain theories of galaxy evolution and cosmological models \citep{bahcall}.
High redshift radio galaxies are ideal targets to find the most massive galaxies at a given redshift. Radio galaxies follow a close relation in the Hubble $K-z$ diagram \citep{lilly84}. At $z\gtrsim 1$ they are $\gtrsim 2$ magnitudes brighter than normal galaxies at these redshifts  \citep{deb02}, and can be used to find the most luminous star-forming populations.
Using galaxy evolution models, \citet{rocca}, find that the brightest luminosity limit of the Hubble $K$ diagram for typical powerful radio galaxies, correspond to the most massive elliptical galaxies of $\sim 10^{12} M_{\odot}$.
One of the most successful technique to find high redshift radio galaxies, has been the ultra steep spectrum criterion (USS) \citep{rot94,cha96a,blu98}.
Selecting sources with steep radio spectra ($\alpha \lesssim -1.30; S_\nu \propto \nu^{\alpha}$) increases dramatically the chance of finding $z>2$ radio galaxies \citep{rot94,cha96b,jarvis}.
It has been known for some time that USS radio galaxies share a close relationship with groups and clusters of galaxies \citep{cha96a,Knopp}.
From cross--correlation analysis \citep{yo1}, there is evidence that USS sources at redshift $\sim 1$ are located in cluster environments, comparable to that expected for clusters of Abell richness class 0 \citep{best2000,best2003}.
Recently, the search for distant forming clusters of galaxies (proto--clusters) associated with radio galaxies has made significant progress in our knowledge of the large-scale structure at high redshifts \citep{pente1997,kurk2000,vene2002,mileynat}.
Some high redshift radio galaxies show very close companion objects along the radio axis.
\citet{roet96}, find a statistical excess of optical companion galaxies up to $\sim$ 80 kpc distance along the radio axis of USS radio sources. \citet{pente2001} reported a similar effect, finding faint close companion galaxies along the radio axis.
\citet{croft} found evidence for a filamentary structure associated with the radio galaxy MRC 1138-262 at $z=2.16$, consisting in a number of X--ray sources \citep{pente2002}, Ly$\alpha$ emitters \citep{pente2000}, H$\alpha$ emitters \citep{kurk2004a}
and EROs \citep{kurk2004b} at redshifts close to that of the radio source, aligned with the axis of the radio emission of the central galaxy.

In this paper we analyze galaxy clustering and surrounding galaxy properties around a large sample of USS radio sources by studying an Abell-type measurement $N_{0.5}$. We examine the dependence on the environmental variations and radio and/or IR source properties, such as the radio size, position angle of the radio structure and central concentration indices and the evolution with redshift.
This paper is organized as follows: Section 2 describes the sample of analyzed, the source extraction, the completeness of the sample and photometry techniques. We
quantify galaxy excess associated to USS radio galaxies in Section 3. We investigate the statistics of aligned companion galaxies respect to the radio axis of USS targets in Section 4. In section 5, we investigate the central light concentration of USS targets. Finally we discuss our results in Section 6.

Unless otherwise stated, we will use a flat cosmology with density parameters $\Omega_{M}=0.3$, $\Omega_{\Lambda}=0.7$ and a Hubble's constant $H_{0}=100$ $h$ kms$^{-1}$ Mpc$^{-1}$.

\setcounter{figure}{0}
\begin{figure*}
\begin{tabular}{ll}
\psfig{file=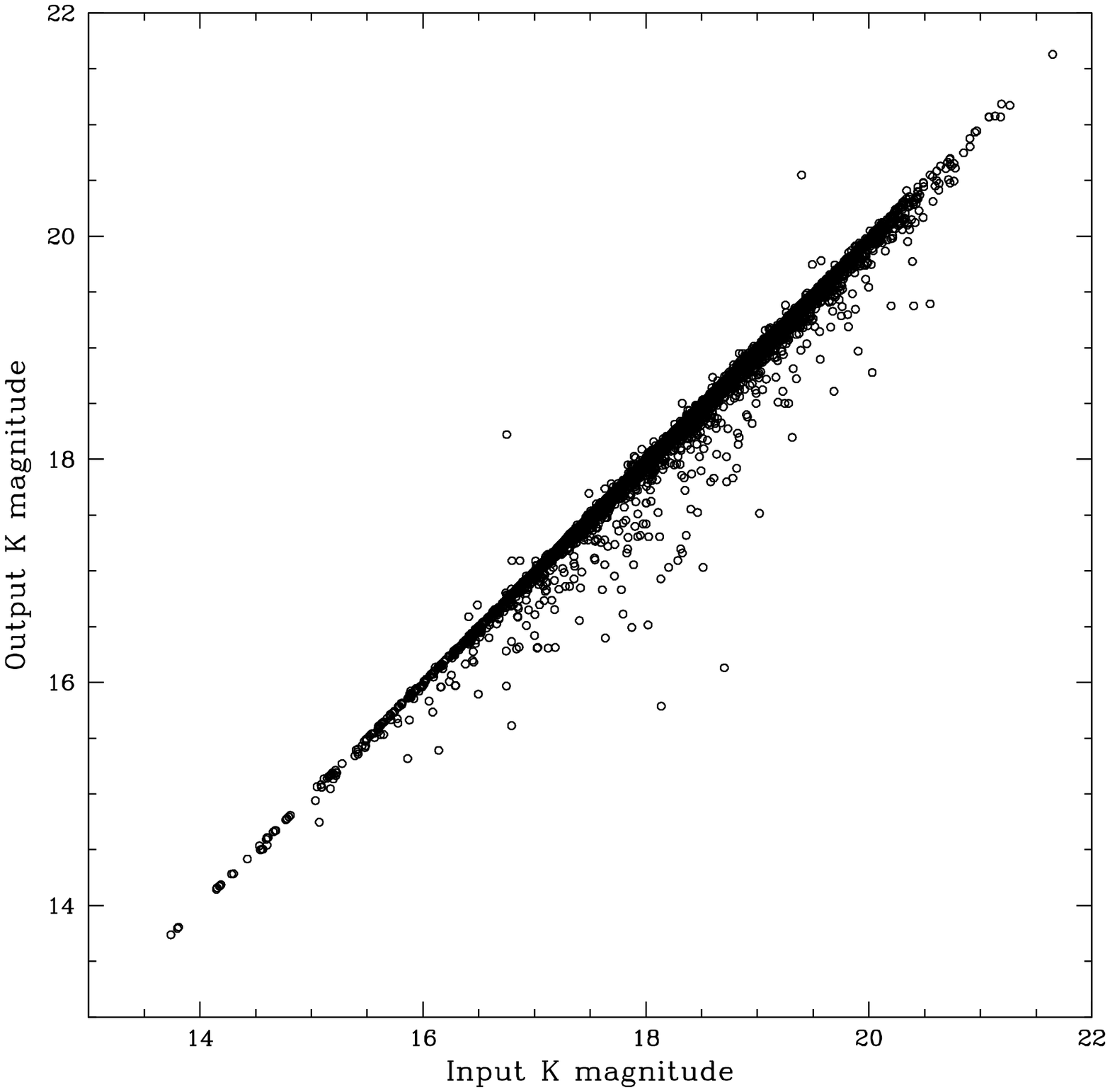,width=8cm}
\psfig{file=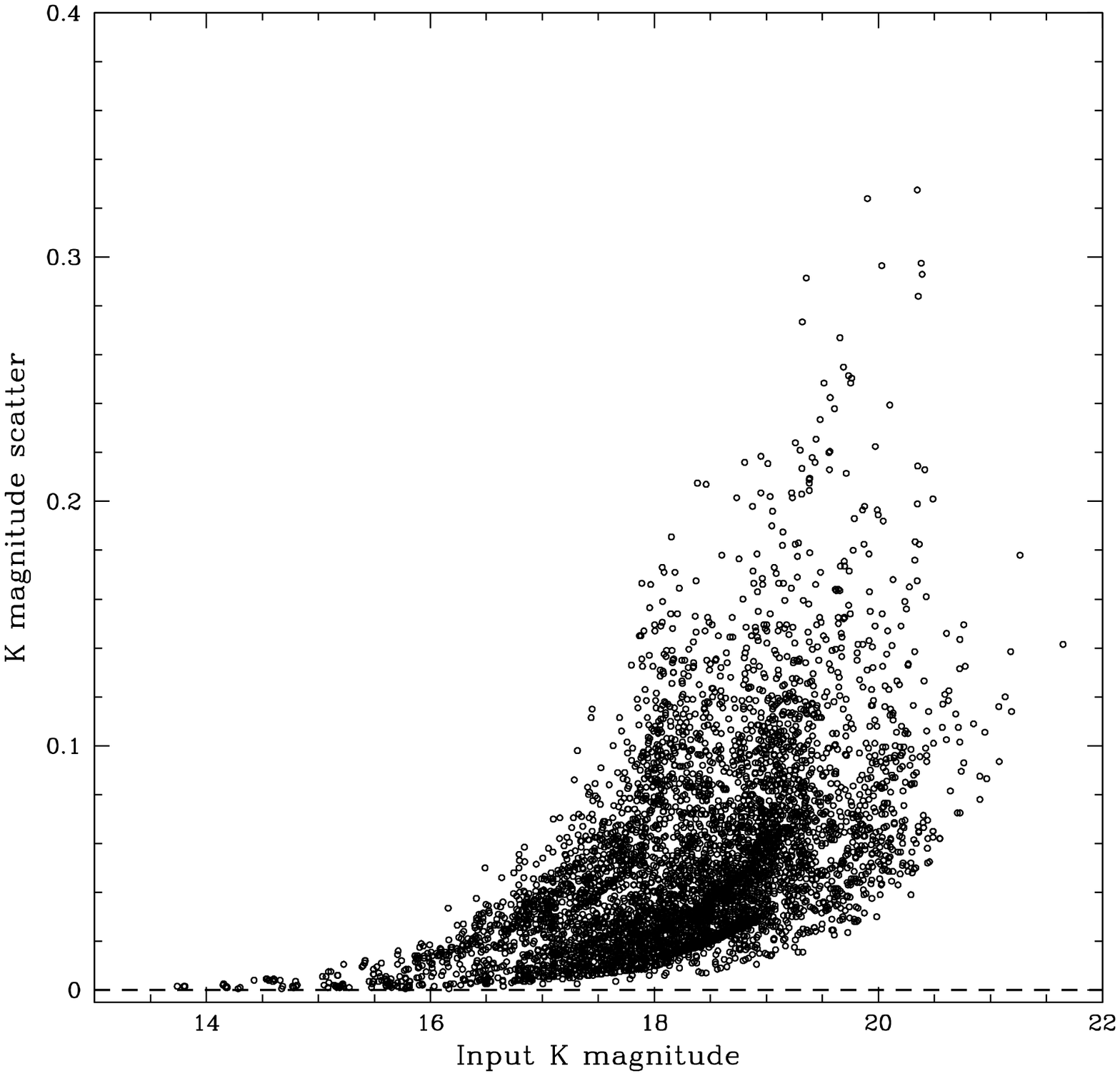,width=8cm}
\end{tabular}
\caption{ The distribution of measured magnitudes (output) as a function of input magnitudes (Left panel). RMS scatter as a function of input magnitudes (Right panel), for the AAT images.}
\newpage
\label{MCtest0}
\end{figure*}

\section{Data acquisition}

The USS sample selection and radio data used for this analysis was presented and described by \citet{sumss}. In short, we used the pre-release version of the SUMSS catalogue and the version 39 of the NVSS catalogue to construct and select a sample of Ultra Steep Spectrum sources. 
We used the Australia Telescope Compact array (ATCA) to measure radio morphologies and accurate radio positions. 
We used 70 radio sources restricted to have an ultra steep radio continuum spectrum, with 53 sources with a spectral index cutoff $\alpha_{843}^{1400}<-1.3$ and 17 sources with $\alpha_{843}^{1400}>-1.3$. We have decided to retain these 17 sources in order to search correlations with other radio and/or IR properties.  
The near-IR data used in this work were obtained using the {\tt IRIS2} instrument at the 3.9m Anglo-Australian Telescope at Siding Spring Observatory. Conditions were mostly photometric with FWHM $\sim 2\farcs$ A $K_{s}$ filter was used. 
The pixel size was 0\farcs446/pixel, resulting in a respective $\sim 8'\times8'$ field of view.
For the 20 sources not detected in the AAT/IRIS2 images, we obtained deeper $K_s$--band images using the Son of Isaac (Sofi) instrument at the ESO 3.5m New Technology Telescope (NTT).
Conditions were photometric with $\sim$0\farcs7 seeing.
The pixel size was 0\farcs292/pixel, resulting in a respective $\sim 5'\times5'$ field of view.
The galaxy sample used in this work consist in galaxies identified in 50 images obtained with the AAT telescope and 20 deeper USS fields obtained with NTT telescope. 
A full description of sample selection, observations and image reductions is given in \cite{sumss}.

\setcounter{figure}{1}
\begin{figure*}
\begin{tabular}{ll}
\psfig{file=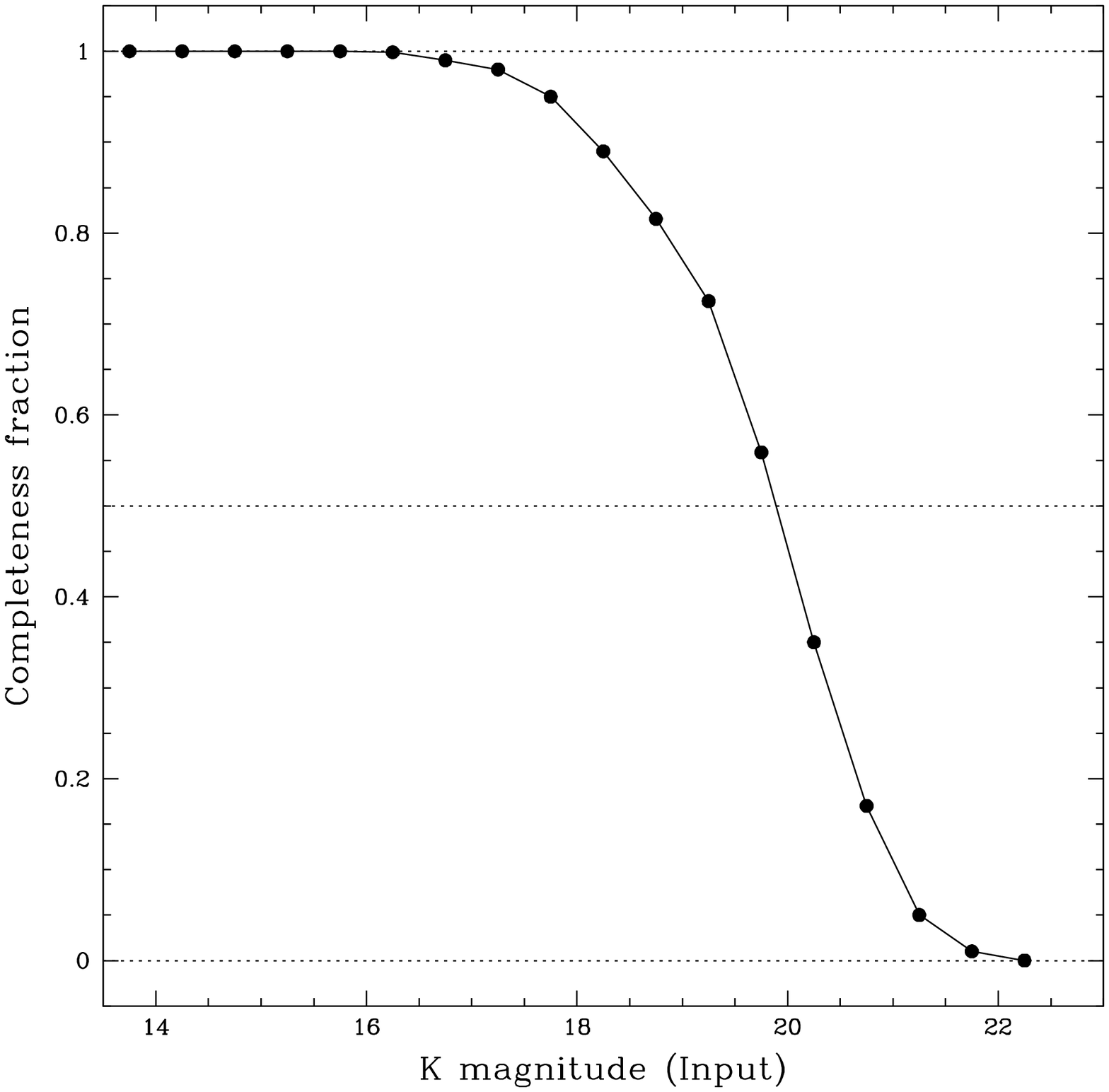,width=8cm}
\psfig{file=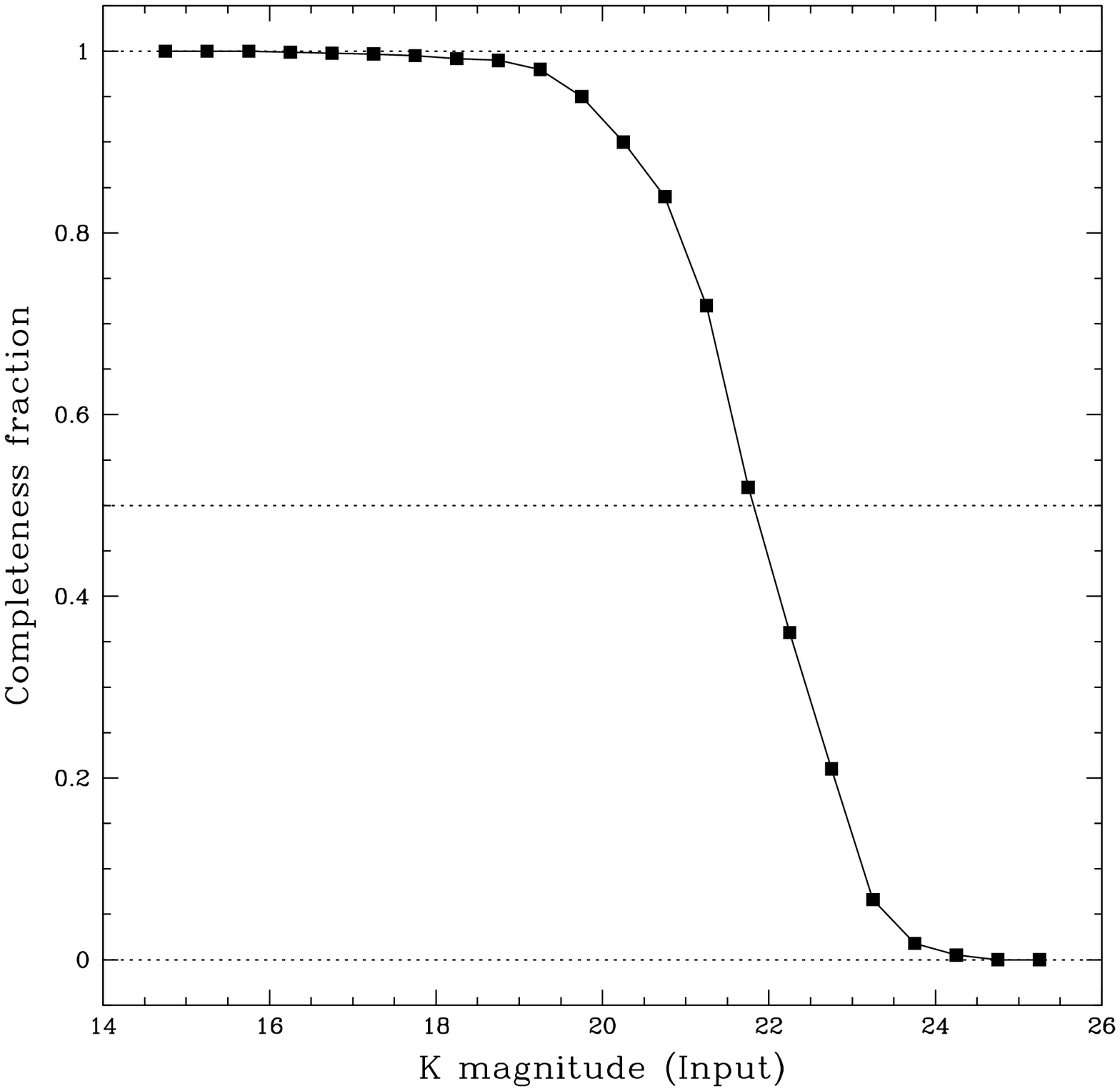,width=8cm}
\end{tabular}
\caption{The completeness fraction for model galaxies for the AAT and NTT images (left and right panels, respectively) as a function of input  magnitude, averaged over 20 fields.  The mean 50\% completeness limits is $K\sim20$ for AAT images and $K\sim21.5$ for the NTT sample.}
\label{MCtest2}
\end{figure*}

\subsection{Photometry and Catalogue Creation}

For object detection and photometry we used the v$2.2.2$ of the SExtractor software package \citep{bertin96}.
For the AAT and NTT images, the source extraction parameters were set such that, to be detected, an object must have a flux in excess of $1.5$ times the local background noise level over at least $N$ connected pixels, according to the seeing conditions ($\sim$5 and $\sim$8 pixels for AAT and NTT images, respectively). Four and six USS sources weren't detected with this detection threshold in the ATT and NTT images, respectively.

SExtractor's {\tt MAG\_BEST} estimator was used to determine the magnitudes
of the sources.
In this work we choose all objects (galaxies) with stellaricity
index $<0.8$, for the AAT images and $<0.9$ for the NTT images.

The result of the detection process was inspected visually in order to ensure that no obvious objects were missed, and that no false detections were entered into the catalogues. Saturated objects and objects lying in the image boundaries were rejected from the catalogues.

Our sample sources is listed in Table~1 and Table~2 in IAU J2000 format, together with the spectral index $\alpha^{1400}_{843}$, position angle of the radio structure determined from the ATCA maps, largest radio size (LAS), the $K_{s}$ counterpart magnitude, the Hill \& Lilly quantity $N_{0.5}$,  the corrected central concentration, the spectroscopic redshift \citep{debspec} and the expected redshift, estimated from the $K-z$ Hubble Diagram and the 1.5 $\sigma$ limiting magnitude per field.

\subsection{Monte-Carlo test of completeness}

In order to investigate the accuracy of the total magnitudes, the complete level of the source extraction (deblending) and set the appropriate value of the $K_s$--magnitude limit for inclusion in our sample, we have produced a Monte-Carlo test.

A series of model galaxies and stars was made using the IRAF\footnote{Image Reduction and Analysis Facility (IRAF), a software system distributed by the National Optical Astronomy Observatories (NOAO).}  ARTDATA {\tt mkobjects} task with typical magnitudes, sizes, redshifts and  Poisson noise, 
similar than our images. We created 20 artificial images with 30\% of the 
galaxies represented by a de Vaucouleurs's law and 70\% with an exponential disk surface  brightness law. 
The model objects were then convolved with a Gaussian function representing the PSF effects. SExtractor was run on the original and to the convolved model images to determine the differences between input and observed due to the effects before mentioned. 
In Figure~\ref{MCtest0}, we plot the input and measured magnitudes of each detected object in the simulations (Left panel) and the scatter of the measured magnitudes as a function of input magnitudes (Right panel).
 We find low scatter generally at at bright magnitudes. At faint magnitudes we find some sources with a large deviations. These are caused by the proximity of a model object to a bright close source. We have checked the images and large errors such as these are avoided in the final output SExtractor catalogues.  

The completeness fraction for model galaxies as a function of input magnitudes is shown in Figure~\ref{MCtest2} (Left panel) for the AAT images. These simulations have shown that the $K_s$--magnitude limit is $K_s\sim20$. We have done similar analysis for the NTT images where the magnitude limit was $K_s\sim21.5$ (See Figure~\ref{MCtest2}, Right panel).

\setcounter{figure}{2}
\begin{figure}
\begin{tabular}{ll}
\psfig{file=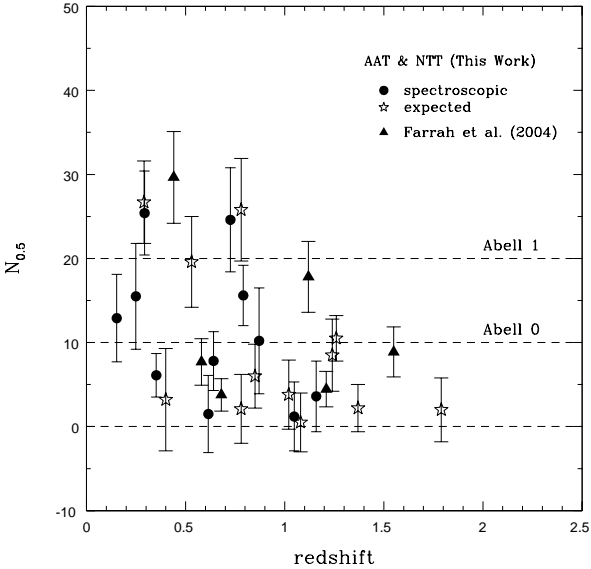,width=8cm}
\end{tabular}
\caption{ Values of the Hill \& Lilly \citep{hill} quantity $N_{0.5}$ vs. redshift for the AAT and NTT sources.
Solid circles represent data with spectroscopic redshift and stars represent data with expected redshift estimated from the $K-z$ Hubble Diagram \citep{sumss}.
Solid triangles represents measurements from Hyperluminous Infrared Galaxies (HLIRGs) taken from \citet{farrah}. 
The horizontal dashed lines represent the conversion between $N_{0.5}$ and the traditional measure of Abell classes quoted in \citet{farrah}}
\label{Nz}
\end{figure}

\section{Cluster richness associated to USS}

In order to obtain an indication of the cluster richnesses associated to USS targets, we have calculated the Hill \& Lilly \citep{hill} quantity. 
This is an Abell--type measurement, defined as the number of excess galaxies within a circle of of projected radius 0.5 Mpc (at the USS redshift), with magnitudes in the range $m_1$ to $m_1+3$, where $m_1$ is the typical magnitude of a USS radio galaxy. 
For comparison with previous studies we adopt a cosmolgy with $\Omega_{M}=0.3$, $\Omega_{\Lambda}=0.7$ and $H_{0}=100$ $h$ kms$^{-1}$ Mpc$^{-1}$, with $h=0.7$.

The average number of galaxies within the same magnitude range in the background was subtracted from the counts in the USS fields, calculated from an annular area region of the image at projected distances greater than $\sim 1$ Mpc distant from the USS targets. The uncertainties on $N_{0.5}$ are dominated by the uncertainties in the background subtraction. We calculated the error on $N_{0.5}$ from the $\sqrt N$ of this number.
In Figure~\ref{Nz}, we plot $N_{0.5}$ vs spectroscopic and expected redshift for USS images with limiting magnitudes fainter than $m_1+3$. We calculated the expected redshift for a USS radio source from the $K-z$ Hubble Diagram \citep{deb02}, using a photometry of 64 kpc radio--galaxy magnitudes. To calculate the 64 kpc metric apertures, we used the average correction for $z>1$,  $K_{64 kpc}=K(8'')+0.2$ (See for instance \citet{sumss}).

Some USS targets analyzed in this work show spectral signatures of QSOs.
In this case, light from the AGN will dominate light from the host galaxy, making a direct measurement of $m_1$ impossible, $m_1$ is therefore estimated from the  $K-z$ Hubble Diagram \citep{deb02,sumss}.
We compare our determinations with the richness measurements of Hyperluminous Infrared Galaxies (HLIRGs) taken from \citet{farrah} and we
also plot the normalization between the value of $N_{0.5}$ and the traditional Abell richness as quoted in this work.

We find a variety of galaxy environments around our sample of USS targets.
Most of the fields studied are compatible with being Abell class 0 richness clusters. 

\setcounter{figure}{3}
\begin{figure*}
\begin{tabular}{ll}
\psfig{file=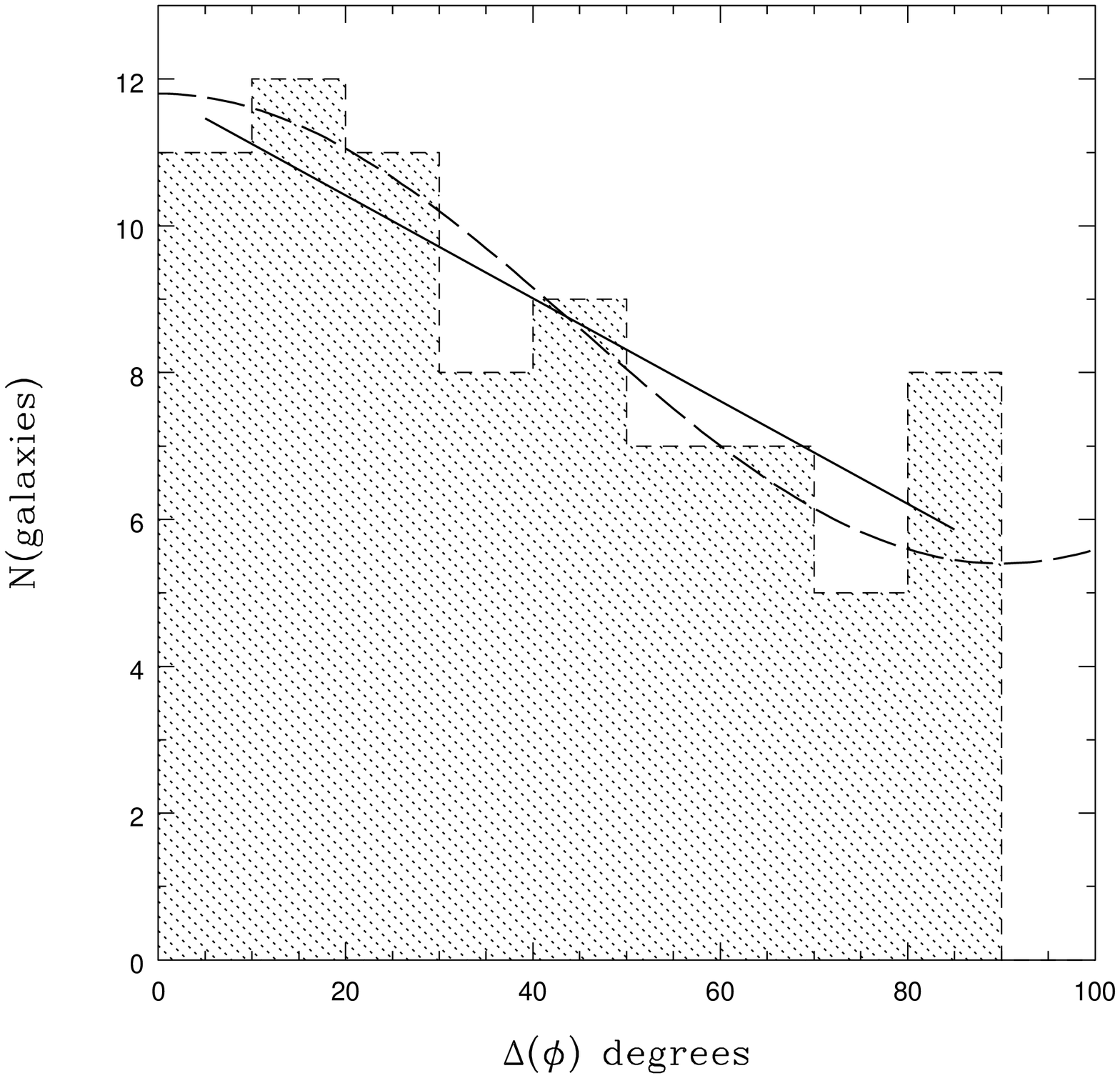,width=8cm}
\psfig{file=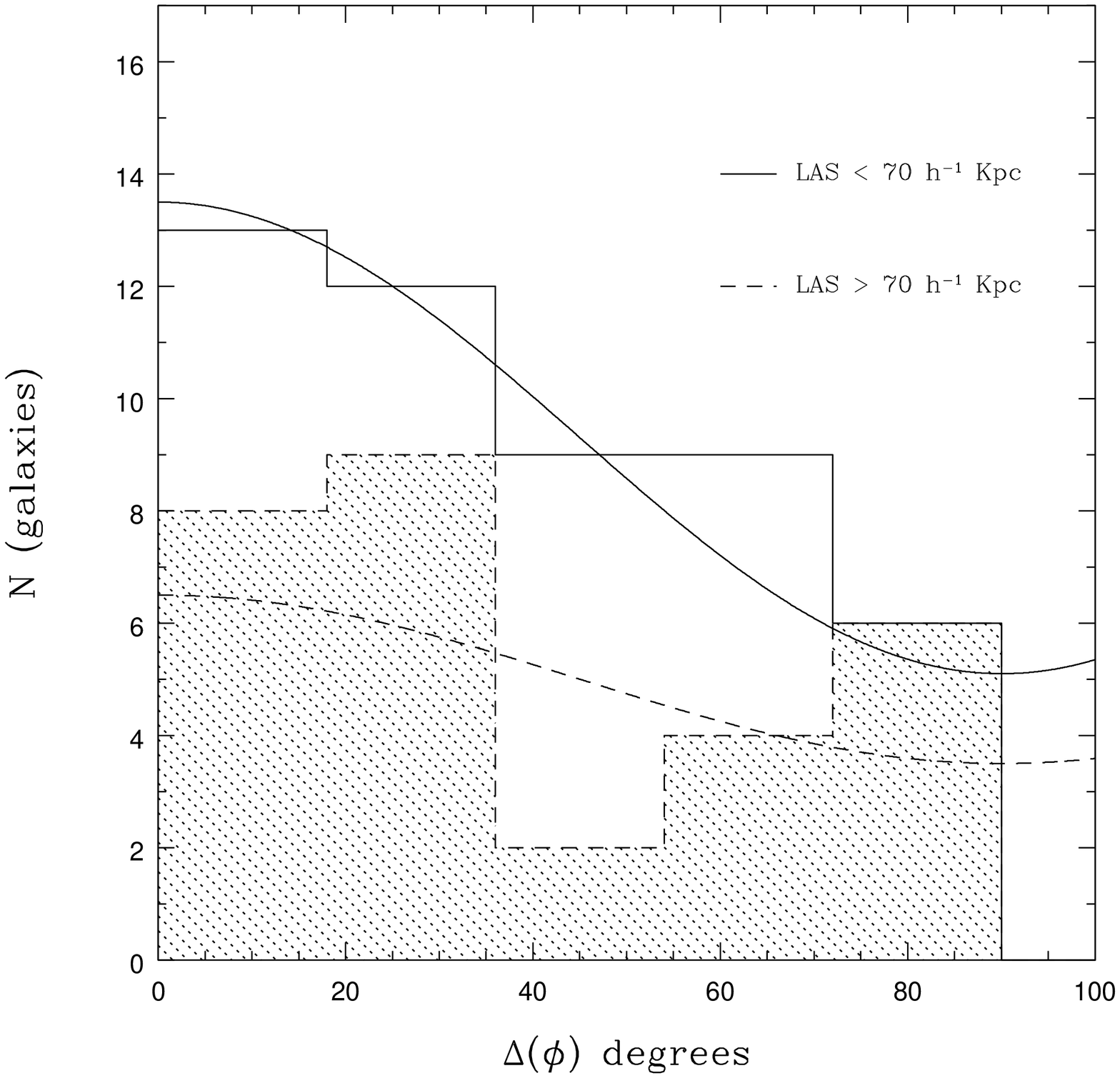,width=8cm}
\end{tabular}
\caption{Left panel: distribution of the orientation of companion galaxies with $15<K_s<20$ within 70 $h^{-1}$ kpc of the USS targets, with respect to the radio axis. The solid line represent the linear fit relation and dashed lines is the double cosine function. Right panel: distribution of the orientation of companion galaxies with $15 <K_s<20$ within 70 $h^{-1}$ kpc of the USS targets with LAS $> 70$ $h^{-1}$ kpc (solid lines histogram) and LAS $< 70$ $h^{-1}$ kpc (grey histogram) in all cases spectroscopic or expected redshift are $z \lesssim 1.5$. }
\label{ang}
\end{figure*}

\section{Distribution of galaxies with respect to the USS radio axis}

\citet{roet96}, find an excess of companion galaxies up to $\sim$ 80 kpc distance along the axes of the radio sources. \citet{pente2001} reported a similar effect, finding nearby faint objects aligned with the direction of the main axis of the radio sources. 
All of these studies only deal with a few ($<$ 30) companion galaxies.
In this section we investigate the distribution of galaxies with respect to the USS radio axis.

In Figure~\ref{ang} (Left panel) we show the distribution of the relative angle between the radius vector 
to companion galaxies with respect to the radio axis for galaxies with $15 <K_s<20$ within 70 $h^{-1}$ kpc of the USS targets, with spectroscopic or expected redshift $z \lesssim 1.5$. 
The distribution shows that the companion galaxies are predominantly located in the direction defined by the radio axis.

Following \citet{sales} we may quantify the departure from an isotropy distribution defining two statistical tools. First, the ratio $f=N_{<30}/N_{>60}$, where $N_{<30}$ and $N_{>60}$ is the number of objects with $\Delta \phi < 30\degr$ and with $\Delta \phi > 60\degr$, respectively. 
For isotropy $f=1$. And the second useful statistical measures of this anisotropy effect can be obtained by fitting a function to the $\phi$ distribution.
 We have adopted a double cosine function 
$N=Acos(2\phi)+B$, and a linear fit  $N=a_{lin}\phi+b$
where A and $a_{lin}$ are the anisotropy parameter in each fit. 
For an isotropic distribution on $\Delta\phi$ we expect $A=a_{lin}=0$.
The resulting value is $f=1.7\pm0.12$.
Uncertainties were calculated based on 20 bootstrap re--sampling of the data.
A suitable estimate of the significance is obtained by calculating $f$ for a random distribution of USS position angles. From this test, we estimate that our results are significant at the 95 \% confidence. 

For the second test, the resulting values are $A=3.2\pm0.6$ and $a_{lin}=-0.07\pm0.02$.
We do not find great differences between the two model fits, however the double cosine function is less sensitive to noise fluctuations at the extremes due to poor number statistics.

In Figure~\ref{ang} (Right panel) we plot the distribution of the orientation of companion galaxies with $15<K_s<20$ within 70 $h^{-1}$ kpc of the USS targets with LAS $> 70$ $h^{-1}$ kpc and with LAS $< 70$ $h^{-1}$ kpc and with spectroscopic or expected redshifts $z \lesssim 1.5$. 
We find that the sample of USS targets with LAS $< 70$ $h^{-1}$ kpc are more aligned with the direction of the main axis of the radio galaxies that those with LAS $> 70$ $h^{-1}$ kpc.  
For the USS sample with LAS $< 70$ $h^{-1}$ kpc we obtain $f=1.8\pm0.15$, $A=4.2\pm0.25$ and $a_{lin}=-0.094\pm0.013$. And for the sample with LAS $> 70$ $h^{-1}$ kpc, $f=1.55\pm0.18$, $A=1.5\pm1.3$ and $a_{lin}=-0.05\pm0.05$ for the first and the second test respectively.

It is worth noting that given that $K$--band traces the evolved stellar
populations (rather than $R$--band which may be
obtain reflection nebulae and young or
star-forming galaxies) our results indicate that
being these evolved populations
aligned with the radio
structure, indicates that the radio orientation
is linked to the large scale structure traced by
galaxies with substantial evolution in these structures.

We can interpret this effect with a dynamical model for the formation of massive galaxies.
\citet{west} proposed a dynamical origin for the alignment effect.
In this picture, powerful high redshift radio galaxies are assumed to represent an early stage 
in the formation of massive galaxies in the centers of rich clusters. 
Galaxy formation proceeds anisotropically by hierarchical merging of smaller protogalactic units 
along preferred directions, which are related to large--scale filamentary features in the surrounding 
mass distribution. As a consequence of this anisotropic formation process, these galaxies are quite prolate in shape.             
Infalling gas will quickly settle into an accretion disc whose angular momentum vector is aligned with the major axis of the galaxy mass distribution.
If the radio jet emerges in the direction of this angular momentum vector, a natural consequence of such formation model is that one would expect companion galaxies to be preferentially located along the radio axis.

The model proposed here does not preclude the possibility that some fraction of the aligned light may have its origin in jet-induced star formation.
In this case, this phenomena is only a second order effect which may act to enhance further the alignments with the radio axis.

\section{Central concentration indices}

In this Section we have analyzed the USS galaxy morphology using an automated classification system based on central concentration of light.
We used the concentration index ($C$), which is the fraction of an object's light contained in the central 30\% of its area as measured in an ellipse aligned with the object and having the same axial ratio (see \citet{abraham94,smail}). 

The interpretation of the observed central light concentration indices has been assessed using galaxy model simulations, which is discussed in the next Subsection. 

\setcounter{figure}{4}
\begin{figure*}
\begin{tabular}{ll}
\psfig{file=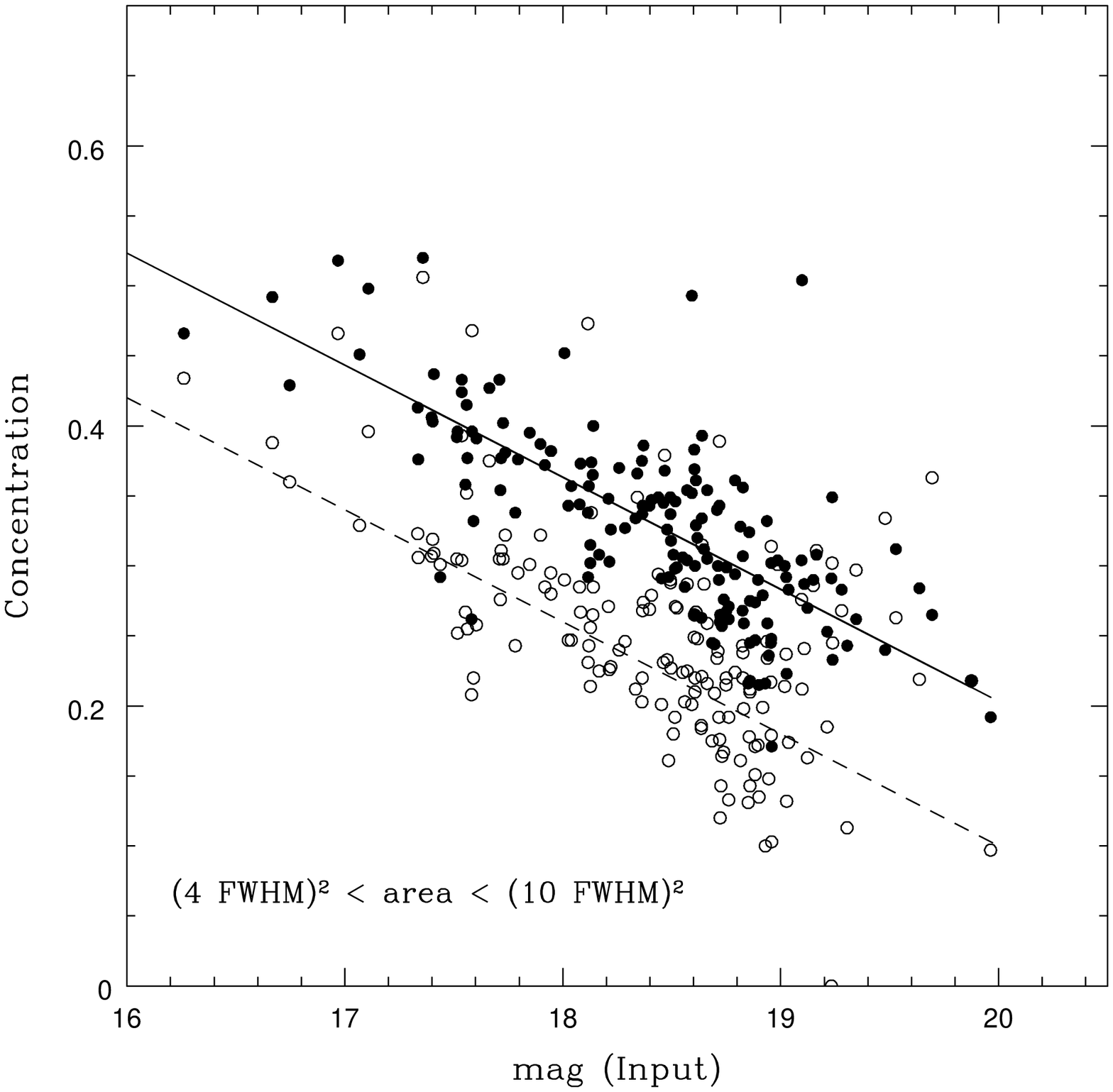,width=8cm}
\psfig{file=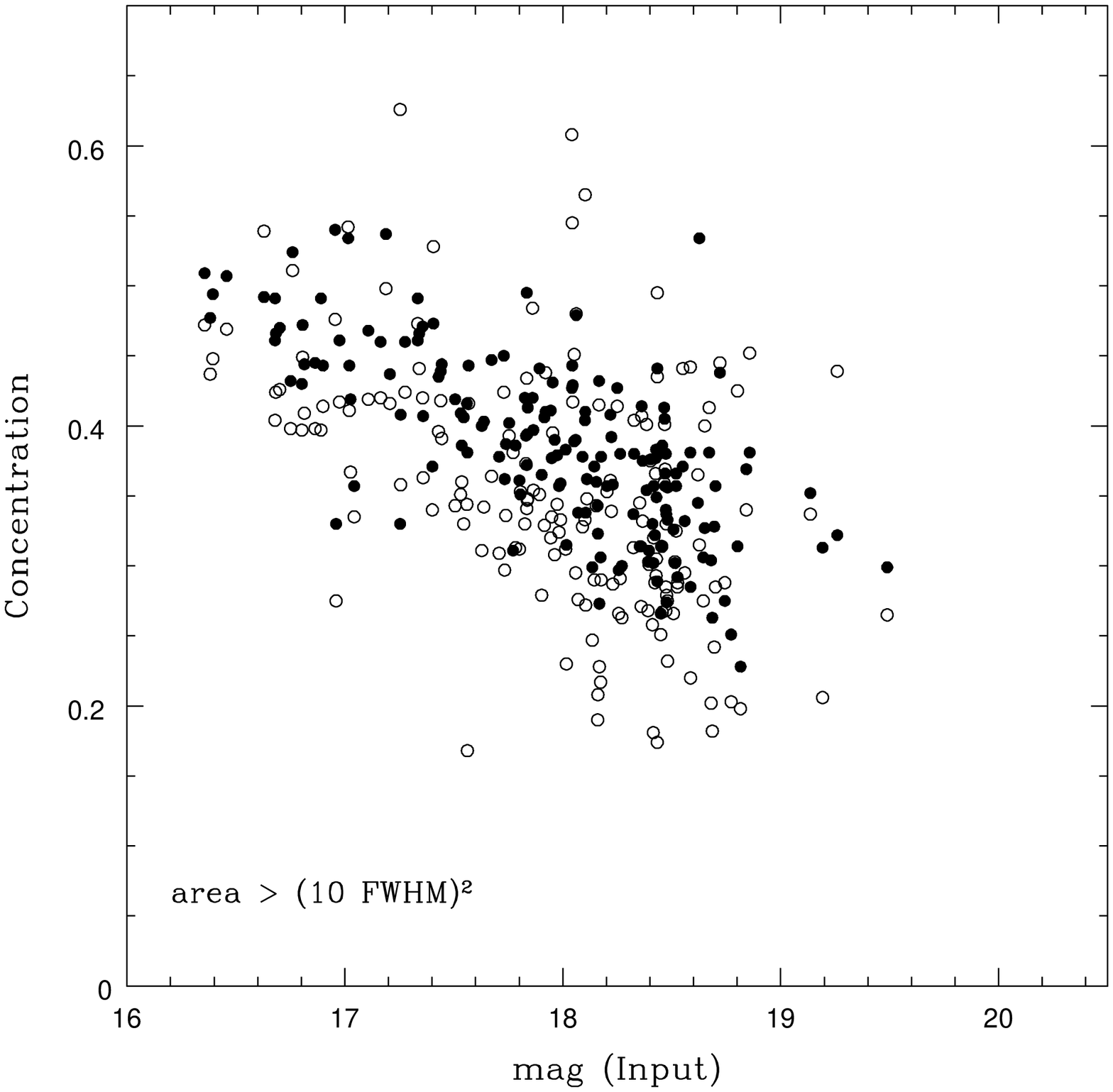,width=8cm}
\end{tabular}
\caption{Left panel: Central concentration indices vs input magnitude from the simulation objects (See for instance \citet{abraham94,abraham96}.  Filled circles and open represent input and output detected objects with (4 PSF FWHM)$^2 < area < $(10 PSF FWHM)$^2$ . The lines represent the mean recovered position of objects due to the effects of seeing degradation. Right panel represent objects with $area > $(10 PSF FWHM)$^2$}
\label{Cmag}
\end{figure*}

\subsection{Simulations and the effects of seeing}
We have done a set of galaxy image simulations in order to investigate the seeing effects degradation on central concentration indices.
We created a series of model galaxies using the IRAF ARTDATA {\tt mkobjects} task, similar to those analyzed in Section 2.4. 
The model objects were then convolved with a Gaussian function representing the PSF effects. SExtractor was run on the original and to the convolved model images to determine the differences between input and output values of $C$.
We have done a similar analysis presented in \citet{abraham94}. We plot the central concentration indices as a function of input magnitude for objects with  (4 PSF FWHM)$^2 < area < $(10 PSF FWHM)$^2$  and those with $> $(10 PSF FWHM)$^2$ (See Figure~\ref{Cmag}). 
We note that small galaxies are largely affected by seeing. The lines represents the best fits between the input and recover positions on the diagram (Figure~\ref{Cmag}, left panel).
On the other hand, we find that the PSF effect not affect the values of $C$ when the projected area is substantially larger than the stellar seeing disk (Figure~\ref{Cmag}, right panel).
Galaxies with projected areas smaller than (4 PSF FWHM)$^2$ were too strongly affected by seeing effects to be usefully classified.

\subsection{Central Concentration indices properties}

In Figure~\ref{Cz} we plot the corrected central concentration indices ($C_{1}$) vs expected redshift or spectroscopic redshift for the USS sample, assuming a correction $\Delta$$C=0.1$, for galaxies with (4 PSF FWHM)$^2 < area < $(10 PSF FWHM)$^2$, where  $\Delta$C is the difference between input and recover positions.
We compare our determinations with the $C_{1}$ vs $z$ values of X--ray AGN selected sources in the Phoenix Deep Survey \citep{hopkins,phoenix} and with determination obtain in the NICMOS Ultra Deep Field (UDF)\citep{gwyn}.
We use public $K'$--band images from Phoenix Deep Survey\footnote{ \scriptsize Data and further
information available at
\texttt{http://www.atnf.csiro.au/people/ahopkins/phoenix/}}, we ran SExtractor with same parameters explained in Section~2.1. We identify the X--ray sources with available spectroscopic or photometric redshifts taken from \citet{phoenix}.
We have done the same analysis with the NICMOS (UDF) images \footnote{\scriptsize Data available at {\texttt{http://orca.phys.uvic.ca/~gwyn/MMM/nicmos.html}}} and then we cross-correlated our SExtractor catalogue with the public photometric redshift catalogue available in the NICMOS UDF web site. 
We have corrected the observed concentration indices as the same way as Section~5.1. 

For this computation we don't measure the concentration indices of the QSOs in the sample, because they are dominated by the 
AGN light, making a direct measurement of $C_1$ impossible.

We find that at $z < 2$ our $C_{1}$ indices are similar with the concentration indices found in the field galaxies and X--ray selected sources. At $z > 2$ the USS sources are different from the general field population. They are highly concentrated in comparison with field galaxies at similar redshifts.

We believe that our USS sample represents the most massive star-forming systems, for a number of reasons.
First, the concentration indices correlates with stellar mass \citep{conse}.
Second, as reviewed in the introduction, at  $z\gtrsim 1$ high redshift radio galaxies are $\gtrsim 2$ magnitudes brighter than normal galaxies at similar redshifts.  This difference suggest that radio galaxies pinpoint the most massive systems out to the highest redshift. 
Third, a recently finished VLT large project \citep{kurk2000,vene2002,vene2004,vene2005,mileynat} has found overdensities of $>$ 20 spectroscopically Ly$\alpha$ and/or H$\alpha$ companion galaxies, associated with galaxy overdensities in all 5 targets studied.\\ 

We compare our richnesses measurements with concentration indices.
In Figure~\ref{NC} we plot the Hill \& Lilly quantity $N_{0.5}$ vs central concentration indices $C_{1}$.
We detect a weak, but significant, positive correlation between $N_{0.5}$ quantity and $C_{1}$.
The Spearman rank correlation coefficients for this relations are $r=0.65$ and $r=0.71$ for the sample with spectroscopic and expected redshifts, respectively.
For the whole redshift sample (spectroscopic and expected) we find $r=0.68$. 

A positive correlation between USS $C_1$ parameter and the environment $N_{0.5}$ value is expected given that the most massive USS hosts (those with larger $C_1$ values)
are likely to be found in the densest galaxy environments.

\setcounter{figure}{5}
\begin{figure}
\begin{tabular}{ll}
\psfig{file=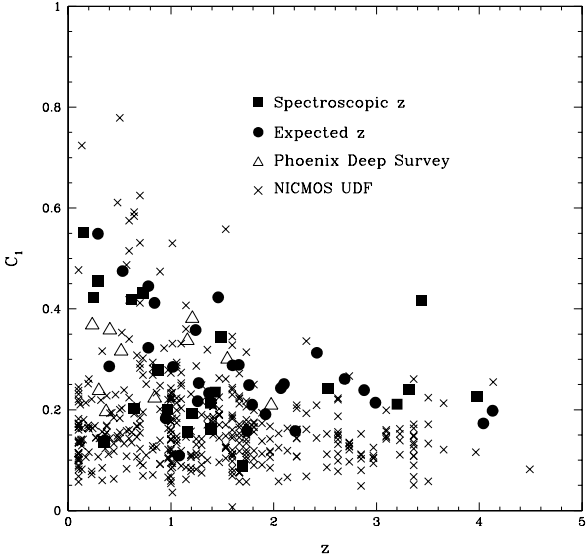,width=8cm}
\end{tabular}
\caption{Corrected central concentration indices vs redshift measured for the AAT/NTT detected objects. Open squares represent data with spectroscopic redshift and filled circles represent data with expected redshift. Crosses are infrared photometric data taken from NICMOS  Ultra Deep Field in the $K'$--band  \citep{gwyn}. Open triangles are data taken from Phoenix Deep Field \citep{phoenix}. }
\label{Cz}
\end{figure}

\setcounter{figure}{6}
\begin{figure}
\begin{tabular}{ll}
\psfig{file=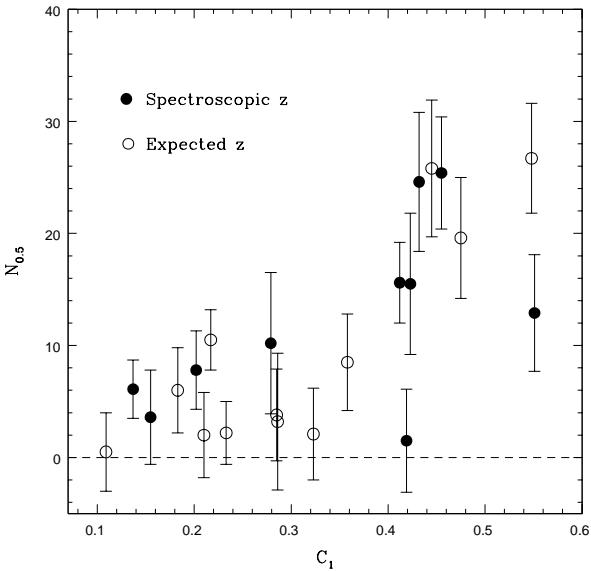,width=8cm}
\end{tabular}
\caption{$N_{0.5}$ vs central concentration indices. Filled circles represent data with spectroscopic redshift and open circles represent data with expected redshift.}
\label{NC}
\end{figure}

\section{Conclusions}

We have analyzed different data sets corresponding to a sample of 70 
Ultra Steep Spectrum (USS) radio sources selected from the 843 MHz Sydney University Molonglo sky Survey (SUMSS) and 1.4 GHz NRAO VLA Sky Survey (NVSS).

We have quantified galaxy excess around USS targets using an Abell--type measurement.
We find that most of the USS fields studied are compatible with being Abell class 0 richness clusters.

A search for companion galaxies along the radio axis present a statistically significant result.
The distribution of companion galaxies around USS radio sources show a 
pronounced tendency for such objects to be found in the direction defined by the radio axis, 
suggesting that they may be related to the presence of the USS radio sources.
We interpret this effect with a model for the formation of radio galaxies via highly anisotropic mergers \citep{west}.
Maybe some fraction of the aligned galaxies may have its origin in jet-induced star formation.

We have also measure the central concentration of light of the USS sample and compare these to the values obtained for field galaxies and X--ray selected galaxies.
We compare our richnesses determinations with concentration indices, and detected a weak, but significant, correlation between $N_{0.5}$ and $C_{1}$ values.
Also, we find that the higher redshift USS radio sources are more concentrated than field galaxies at similar redshifts.
This difference suggest that USS radio galaxies trace the massive systems at high redshift.

\section{Acknowledgements}

We thank the Referee for helping us to improve the original version of this work. 
Carlos Bornancini is specially grateful to Julian Mart\'{\i}nez for helpful comments and suggestions.
This work was partially supported by the
Consejo Nacional de Investigaciones Cient\'{\i}ficas y T\'ecnicas,
Agencia de Promoci\'on de Ciencia y Tecnolog\'{\i}a,  Fundaci\'on Antorchas
and Secretaria de Ciencia y
T\'ecnica de la Universidad Nacional de C\'ordoba.

{}

\setcounter{table}{0}
\begin{onecolumn}
\label{t3}
\small
Table 1. NTT Sample characteristics. Designation in IAU J2000 format, J2000 coordinates of the $K-$band identification, the spectral index $\alpha_{843}^{1400}$, position angle of the radio structure determined from the ATCA maps, largest radio size, the $K$ counterpart magnitude, the Hill \& Lilly quantity $N_{0.5}$, the corrected central concentration, the spectroscopic and the expected redshift and the 1.5 $\sigma$ limiting magnitude per field.
\begin{center}
\begin{tabular}{lcrrcccccc}
\hline
(1) &(2)&(3)&(4)&(5)&(6)&(7)&(8)&(9)&(10)\\
Name & $\alpha_{843}^{1400}$& PA & LAS & K$-$mag. &$N_{0.5}$& $C_{1}$& $z$ &$z$ &Limiting mag.\\
& &\degr & {\scriptsize $h^{-1}$kpc} & $_{\tt MAG\_BEST}$ &  & &{\scriptsize spectroscopic} &{\scriptsize expected}&{\scriptsize 1.5$\sigma$}\\
\hline
NVSS~J002738$-$323501         &-1.59$\pm$0.20  &23      &269.1  & 18.06$\pm$0.06    &2.0$\pm$3.8   &0.210   &\nodata                &1.79    & 21.4\\ 
NVSS~J011606$-$331241         &-1.70$\pm$0.20  &\nodata &$<$20.7& 18.27$\pm$0.06    &6.1$\pm$7.6   &0.137   &0.352$\pm$0.001        &1.50    & 21.4\\ 
NVSS~J014413$-$330457         &-0.82$\pm$0.27  &97      &228.6  & 18.55$\pm$0.08    &\nodata       &0.239   &\nodata                &2.88    & 21.5\\
NVSS~J015436$-$333425         &-1.59$\pm$0.12  &\nodata &$<$35.5& 19.67$\pm$0.12    &\nodata       &\nodata &\nodata                &1.74    & 21.1\\
NVSS~J202945$-$344812         &-1.25$\pm$0.10  &167     &112.0  & 19.42$\pm$0.02    &\nodata       &\nodata &1.497$\pm$0.002$^{b}$  &1.14    & 21.4\\ 
NVSS~J204420$-$334948$^{a}$   &-1.60$\pm$0.11  &15      &\nodata&\nodata            &\nodata       &\nodata &\nodata                &\nodata & 21.3\\ 
NVSS~J230035$-$363410         &-1.67$\pm$0.14  &\nodata &$<$27.0& 20.50$\pm$0.22    &\nodata       &0.173   &\nodata                &4.04    & 21.1\\
NVSS~J230123$-$364656         &-1.61$\pm$0.14  &\nodata &$<$31.7& 19.63$\pm$0.14    &\nodata       &0.211   &3.220$\pm$0.002        &2.62    & 21.5\\ 
NVSS~J230527$-$360534$^{a}$   &-1.64$\pm$0.10  &\nodata &\nodata&\nodata            &\nodata       &\nodata &\nodata                &\nodata & 21.1\\ 
NVSS~J230954$-$365653         &-1.15$\pm$0.18  &25      &\nodata & 19.75$\pm$0.15   &\nodata       &0.198   &\nodata                &6.27  &21.1 \\ 
NVSS~J231144$-$362215         &-1.32$\pm$0.12  &69      &92.4    & 20.20$\pm$0.17   &\nodata       &0.243   &2.531$\pm$0.002        &4.45    &21.6\\
NVSS~J231338$-$362708         &-1.50$\pm$0.12  &123     &10.7    & 19.10$\pm$0.10   &\nodata       &0.089   &1.838$\pm$0.002        &2.93    &21.4\\
NVSS~J231726$-$371443         &-1.23$\pm$0.09  &19      &80.3    & 19.07$\pm$0.10   &\nodata       &\nodata &\nodata                &1.99    &21.2\\ 
NVSS~J231727$-$352606$^{a}$   &-1.33$\pm$0.09  &124     &21.2    & \nodata          &\nodata       &\nodata &3.874$\pm$0.002        &\nodata &21.5\\ 
NVSS~J232001$-$363246         &-2.01$\pm$0.16  &\nodata &$<$28.8 & 19.79$\pm$0.17   &\nodata       &0.198   &1.483$\pm$0.001        &4.13    &21.6\\ 
NVSS~J232100$-$360223         &-1.77$\pm$0.14  &\nodata &31.3    & 20.00$\pm$0.20   &\nodata       &0.240   &3.320$\pm$0.005        &3.96    &21.6\\ 
NVSS~J232219$-$355816$^{a}$   &-1.87$\pm$0.10  &149     &\nodata &\nodata           &\nodata       &\nodata &\nodata                &\nodata &21.5\\ 
NVSS~J233558$-$362236         &-1.24$\pm$0.16  &148     &209.5   & 16.52$\pm$0.01   &15.5$\pm3.6$  &0.412   &0.791$\pm$0.001        &0.76    &21.4\\ 
NVSS~J234137$-$342230$^{a}$   &-1.55$\pm$0.15  &\nodata &\nodata &\nodata           &\nodata       &\nodata &\nodata                &\nodata &21.2\\ 
NVSS~J235137$-$362632$^{a}$   &-1.62$\pm$0.11  &10      &\nodata &\nodata           &\nodata       &\nodata &\nodata                &\nodata &21.0\\ 
\hline
\end{tabular}
\end{center}
$^{a}$ Radio position.\\
$^{b}$ QSO.
\end{onecolumn}

\setcounter{table}{1}
\begin{onecolumn}
\small
Table 2. Same as table 1 for the AAT USS Sample.
\begin{center}
\label{t4}
\begin{tabular}{lcrrcccccc}
\hline
(1) &(2)&(3)&(4)&(5)&(6)&(7)&(8)&(9)&(10)\\
Name &  $\alpha_{843}^{1400}$& PA & LAS & K$-$mag. &$N_{0.5}$& $C_{1}$& $z$ &$z$ &Limiting mag.\\
&& \degr & {\scriptsize $h^{-1}$kpc} & $_{\tt MAG\_BEST}$ &  & &{\scriptsize spectroscopic} &{\scriptsize expected}&{\scriptsize 1.5$\sigma$}\\
\hline
NVSS~J002001$-$333408    &-1.38$\pm$0.13         &38       &71.0   &17.91$\pm$0.08 &\nodata      &0.423   & continuum$^{a}$    &1.46& 20.0 \\ 
NVSS~J002112$-$321208    &-1.66$\pm$0.13         &\nodata  &$<$27.8 &19.25$\pm$0.19 &\nodata      &0.261  &\nodata             &2.69& 20.0 \\  
NVSS~J002131$-$342225    &-1.57$\pm$0.15         &116      &34.2   &15.29$\pm$0.01 &15.5$\pm$6.3 &0.423   &0.249$\pm$0.001     &0.37& 20.0 \\ 
NVSS~J002352$-$332338    &-1.10$\pm$0.13         &79       &221.3  &19.20$\pm$0.25 &\nodata      &\nodata &\nodata             &1.98& 19.8 \\ 
NVSS~J002359$-$325756    &-1.69$\pm$0.09         &94       &12.1   &18.39$\pm$0.12 &\nodata      &0.158   &\nodata             &2.21& 20.0 \\ 
NVSS~J002402$-$325253    &-1.74$\pm$0.10         &4        &22.1   &19.07$\pm$0.23 &\nodata      &\nodata & 2.043$\pm0.002$    &2.57& 20.0 \\ 
NVSS~J002415$-$324102    &-1.78$\pm$0.10         &\nodata  &$<$28.5& \nodata       &\nodata      &\nodata &\nodata             &\nodata& 19.8 \\ 
NVSS~J002427$-$325135    &-1.54$\pm$0.11         &171      &38.9   &17.19$\pm$0.09 &2.2$\pm$2.7  &0.233   &continuum           &1.37& 20.3 \\ 
NVSS~J011032$-$335445    &-1.52$\pm$0.16         &23       &$<$29.6&18.67$\pm$0.16 &\nodata      &0.158   &continuum           &1.74& 20.0 \\ 
NVSS~J011643$-$323415    &-0.93$\pm$0.33         &121      &250.8  &18.05$\pm$0.13 &\nodata      &0.289   &\nodata             &1.66& 20.0 \\ 
NVSS~J014529$-$325915    &-1.32$\pm$0.15         &41       &67.4   &15.80$\pm$0.01 &19.6$\pm$5.4 &0.475   &\nodata             &0.53& 20.0 \\ 
NVSS~J015232$-$333952    &-1.32$\pm$0.09         &7        &52.6   &16.26$\pm$0.02 &1.5$\pm$4.6  &0.419   &0.6148$\pm$0.001    &0.60& 20.5 \\ 
NVSS~J015324$-$334117    &-1.23$\pm$0.15         &91       &49.2   &14.51$\pm$0.01 &12.9$\pm$5.2&0.551   &0.1525$\pm$0.0004   &0.22& 20.2 \\ 
NVSS~J015418$-$330150    &-1.23$\pm$0.15         &\nodata  &$<$29.4&      \nodata  & \nodata     &\nodata &\nodata             &\nodata& 20.5 \\ 
NVSS~J015544$-$330633    &-1.14$\pm$0.11         &115      &105.5  &16.93$\pm$0.05 &1.2$\pm$4.1  &\nodata &1.048$\pm$0.002$^b$ &0.94& 20.0 \\ 
NVSS~J021308$-$322338    &-1.42$\pm$0.11         &\nodata  &26.2   &19.14$\pm$0.22 &\nodata      &0.296   &3.98$\pm$0.001      &3.27& 19.4 \\  
NVSS~J021359$-$321115    &-1.29$\pm$0.15         &\nodata  &$<$35.6&17.89$\pm$0.07 &\nodata      &0.288   &\nodata             &1.60& 20.0 \\ 
NVSS~J021545$-$321047    &-1.59$\pm$0.10         &98       &42.6   &19.20$\pm$0.22 &\nodata      &\nodata &\nodata             &2.44& 20.0 \\ 
NVSS~J021716$-$325121    &-1.52$\pm$0.12         &44       &24.2   &18.75$\pm$0.20 &\nodata      &0.213   &1.384$\pm$0.002     &1.76& 20.0 \\  
NVSS~J030639$-$330432    &-1.70$\pm$0.10         &\nodata  &$<$29.0&17.88$\pm$0.11    &\nodata   &0.193   &1.201$\pm$0.001     &1.42& 20.2 \\ 
NVSS~J202026$-$372823    &-1.34$\pm$0.10         &\nodata  &$<$29.6&18.56$\pm$0.15    &\nodata   &0.235   &1.431$\pm$0.001     &1.67& 20.2 \\ 
NVSS~J202140$-$373942    &-1.34$\pm$0.14         &29.1     &45.8   &15.20$\pm$0.01 &3.2$\pm$6.1  &0.286   &\nodata             &0.40& 19.6 \\ 
NVSS~J202518$-$355834    &-1.55$\pm$0.11         &\nodata  &$<$35.5&18.43$\pm$0.10 &\nodata      &0.249   &\nodata             &1.76& 20.0 \\ 
NVSS~J202856$-$353709    &-1.07$\pm$0.10         &129      &186.6  &16.60$\pm$0.04 &2.1$\pm$4.2  &0.323   &continuum           &0.78& 20.0 \\ 
NVSS~J204147$-$331731    &-1.35$\pm$0.13         &168      &23.1   &16.86$\pm$0.05 &10.2$\pm$6.3 &0.279   &0.871$\pm$0.001     &0.84& 20.1 \\ 
NVSS~J213510$-$333703    &-1.46$\pm$0.11         &15       &$<$20.4&\nodata        &\nodata      &\nodata &2.518$\pm$0.001     &2.11& 19.0 \\ 
NVSS~J225719$-$343954    &-1.69$\pm$0.09         &\nodata  &30.5   &16.53$\pm$0.02 &24.6$\pm$6.2&0.432   &0.726$\pm$0.001     &0.70& 19.8 \\  
NVSS~J230203$-$340932    &-1.31$\pm$0.14         &91       &148.6  &17.34$\pm$0.07 &3.6$\pm$4.2 &0.155   &1.159$\pm$0.001     &2.14& 20.4 \\  
NVSS~J230404$-$372450    &-1.24$\pm$0.12         &126      &181.5  &17.40$\pm$0.10 &0.5$\pm$3.5  &0.109   &continuum           &1.08& 20.4 \\ 
NVSS~J230822$-$325027    &-1.68$\pm$0.11         &141      &9.9    &18.47$\pm$0.10 &\nodata      &0.251   &\nodata             &2.10& 20.0 \\ 
NVSS~J230846$-$334810    &-1.07$\pm$0.10         &114      &110.4  &16.97$\pm$0.02 &25.8$\pm$6.1 &0.445   &continuum           &0.78& 20.1 \\ 
NVSS~J231016$-$363624    &-1.48$\pm$0.13         &3        &85.0   &14.60$\pm$0.05 &26.7$\pm$4.9 &0.549   &\nodata             &0.29& 20.2 \\ 
NVSS~J231229$-$371324    &-1.40$\pm$0.12         &161      &32.1   &17.31$\pm$0.08 &10.5$\pm$2.7 &0.217   &continuum           &1.26&20.4  \\ 
NVSS~J231311$-$361558    &-0.54$\pm$0.14         &48       &367.0  &17.49$\pm$0.08 &8.5$\pm$4.3 &0.358   &\nodata             &1.24&20.5  \\ 
NVSS~J231317$-$352133    &-1.35$\pm$0.13         &102      &33.8   &19.14$\pm$0.25 &\nodata      &\nodata &\nodata             &2.24&20.0  \\ 
NVSS~J231335$-$370609    &-1.21$\pm$0.11         &\nodata  &$<$23.6&18.63$\pm$0.18 & \nodata     &0.191   &\nodata             &1.92&20.0  \\ 
NVSS~J231341$-$372504    &-1.23$\pm$0.18         &155      &250.3  &16.97$\pm$0.07 &6.0$\pm$3.8  &0.183   &continuum           &0.95&20.0  \\ 
NVSS~J231357$-$372413    &-1.58$\pm$0.10         &47      &11.8    &16.00$\pm$0.03 &\nodata      &0.161   &1.393$\pm$0.001     &0.76&20.4\\
NVSS~J231402$-$372925    &-1.37$\pm$0.08         &48       &22.2   &18.69$\pm$0.15 &\nodata      &0.416   &3.450$\pm$0.005     &2.15&20.0   \\ 
NVSS~J231459$-$362859    &-1.16$\pm$0.11         &16       &55.2   &19.15$\pm$0.15 & \nodata     &0.313   &\nodata             &2.42&20.0    \\ 
NVSS~J231519$-$342710    &-1.41$\pm$0.11         &106      &119.8  &18.10$\pm$0.13 & \nodata     &0.200   &0.970$\pm$0.001     &1.41&20.1   \\ 
NVSS~J232014$-$375100    &-1.11$\pm$0.13         &128      &262.1  &17.60$\pm$0.10 &\nodata      &0.253   &\nodata             &1.27&20.0 \\ 
NVSS~J232058$-$365157    &-1.38$\pm$0.09         &\nodata  &29.6   &18.65$\pm$0.12 &\nodata      &0.344   &\nodata             &2.28&20.0  \\ 
NVSS~J232322$-$345250    &-1.19$\pm$0.15         &\nodata  &31.0   &17.07$\pm$0.08 &3.8 $\pm$4.1 &0.285   &continuum           &1.02&20.1  \\ 
NVSS~J232602$-$350321    &-0.94$\pm$0.24         &140      &46.8   &14.23$\pm$0.11 &25.4$\pm$5.0&0.455   &0.293$\pm$0.001     &0.27&19.8 \\ 
NVSS~J232651$-$370909    &-1.56$\pm$0.10         &\nodata  &$<$21.6&19.16$\pm$0.10 &\nodata      &0.214   &\nodata             &2.99&20.2  \\ 
NVSS~J232956$-$374534    &-1.53$\pm$0.10         &\nodata  &$<$35.0&19.06$\pm$0.21 &\nodata      &0.243   & \nodata            &2.07&20.0 \\  
NVSS~J233729$-$355529    &-1.33$\pm$0.09         &72       &43.9   &\nodata        &\nodata      &\nodata &\nodata             &\nodata&20.0\\ 
NVSS~J234145$-$350624    &-1.30$\pm$0.08         &\nodata  &$<$24.0&15.94$\pm$0.04 &7.8$\pm$3.5 &0.202   &0.641$\pm$0.001     &0.60&20.0 \\ 
NVSS~J234904$-$362451    &-1.31$\pm$0.10         &137      &56.3   &17.63$\pm$0.14 &3.8$\pm$2.5  &\nodata & 1.520$\pm$0.003$^b$&1.33&20.5  \\ 
\hline

\end{tabular}
\end{center}
$^{a}$ Objects with continuum features, no emission lines.\\
$^{b}$ QSO.
\end{onecolumn}

\end{document}